\documentclass[a4paper,twocolumn,11pt]{quantumarticle}
\pdfoutput=1
\usepackage{amsmath}
\usepackage{amssymb}
\usepackage{amsthm}
\usepackage{graphicx}
\usepackage{hyperref}
\usepackage[dvipsnames]{xcolor} 
\usepackage{booktabs} 
\usepackage{siunitx}
\usepackage[T1]{fontenc}

\newcommand{\todo}[1]{}

\begin{document}

\title{Differentiable Logical Programming for Quantum Circuit Discovery and Optimization}

\author{Antonin Sulc}
\email{asulc@lbl.gov}
\affiliation{Lawrence Berkeley National Laboratory, Berkeley, California 94720, USA}
\orcid{0000-0001-7767-778X}

\begin{abstract}
Designing high-fidelity quantum circuits remains challenging, and current paradigms often depend on heuristic, fixed-ansatz structures or rule-based compilers that can be suboptimal or lack generality. We introduce a neuro-symbolic framework that reframes quantum circuit design as a differentiable logic programming problem. Our model represents a scaffold of potential quantum gates and parameterized operations as a set of learnable, continuous ``truth values'' or ``switches,'' $s \in [0, 1]^N$. These switches are optimized via standard gradient descent to satisfy a user-defined set of differentiable, logical axioms (e.g., correctness, simplicity, robustness). We provide a theoretical formulation bridging continuous logic (via T-norms) and unitary evolution (via geodesic interpolation), while addressing the barren plateau problem through biased initialization. We illustrate the approach on tasks including discovery of a 4-qubit Quantum Fourier Transform (QFT) from a scaffold of 21 candidate gates. We also report hardware-aware adaptation experiments on the 156-qubit IBM Fez processor, where the method autonomously adapted to both gradual noise drift (24.2~pp over static baseline) and catastrophic hardware failure (46.7~pp post-failure improvement), using only measurement-driven gradient updates with no hardwired bias or prior path preference.
\end{abstract}
\maketitle

\section{Introduction}
\label{sec:introduction}

Quantum computation promises to solve problems considered intractable for classical computers, with applications spanning materials science and quantum chemistry, drug discovery, and complex optimization problems in areas such as finance and logistics~\cite{nielsen_chuang_2010,mcardle_quantum_2020,cao_quantum_2019,orus_quantum_finance_2019,bova_commercial_2021}.
However, realizing this potential depends on our ability to design and execute high-fidelity quantum circuits. In the current noisy intermediate-scale quantum (NISQ) era, this task is constrained; hardware is limited by gate error rates, coherence times, and qubit connectivity, placing a strong emphasis on circuit efficiency and robustness~\cite{preskill_2018}.

The central challenge is that finding an optimal quantum circuit for a given task is a notoriously difficult combinatorial search problem. Current approaches to circuit design, while powerful, face fundamental limitations.

Manual, human-derived circuits for canonical algorithms like the Quantum Fourier Transform (QFT) or Grover's search are the product of significant ingenuity but are not generalizable to novel problems, such as simulating an arbitrary molecular Hamiltonian. Variational Quantum Algorithms (VQAs), the dominant NISQ paradigm~\cite{peruzzo_2014, farhi_2014}, typically rely on a pre-defined, fixed-structure ``ansatz.'' The performance of the entire algorithm is critically sensitive to this heuristic design choice. A poor ansatz can lead to an inability to express the solution state or suffer from barren plateaus, where training gradients vanish exponentially~\cite{mcclean_2018}. While adaptive methods like ADAPT-VQE~\cite{grimsley_2019} build an ansatz iteratively, they employ a greedy search that may not find a globally optimal structure.

\begin{figure*}[t]
    \centering
    \includegraphics[width=1.0\linewidth]{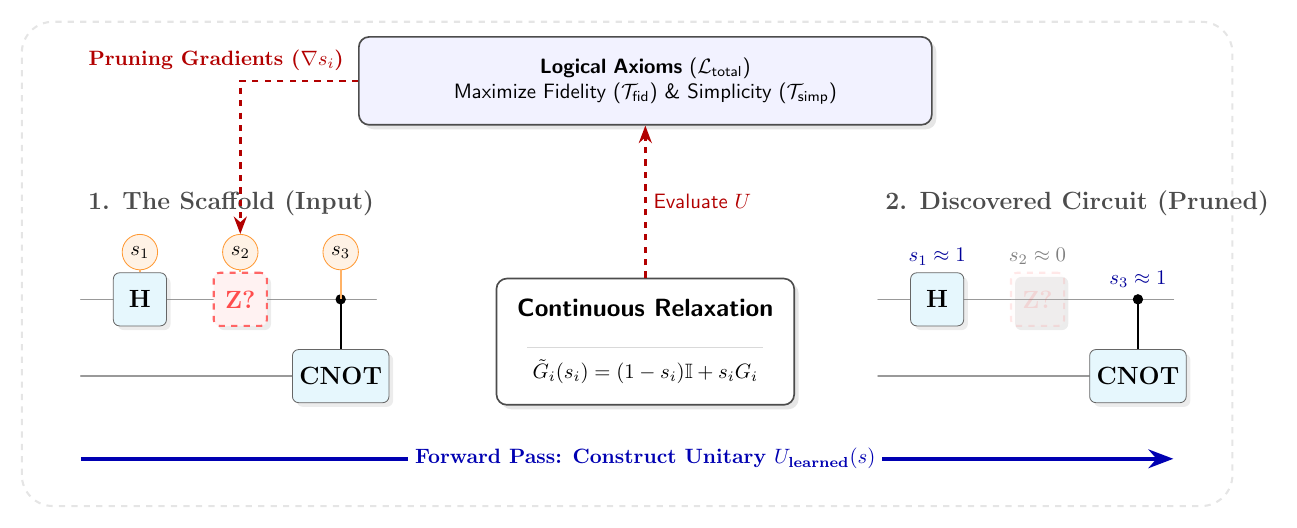}
    
    \caption{\textbf{Conceptual overview of the Differentiable Logical Programming framework for quantum circuit design.} The entire process, from the logical axioms to the circuit structure, is connected by differentiable operations, allowing for end-to-end optimization using standard gradient-based methods. This workflow unifies discrete structural search and continuous parameter optimization.}
    \label{fig:workflow}
\end{figure*}

On the optimization front, quantum compilers~\cite{sivarajah_2020} apply pre-programmed, rule-based graph-rewriting identities (e.g., $H$-$H$ $\to$ $I$). These systems are limited to the set of rules known \textit{a priori}. Other non-differentiable search methods, such as evolutionary algorithms~\cite{evolving_circuits_2002} or SAT solvers for circuit synthesis~\cite{amy_2013_sat,meuli_2019_sat}, must rely on heuristic search strategies and cannot leverage the highly-efficient, gradient-based optimization tools that have revolutionized machine learning.

Reinforcement-learning approaches have also been proposed for quantum architecture search on near-term hardware~\cite{patel2024curriculum, kundu2026reinforcement, kundu2025tensorrl}. These methods improve search efficiency under realistic noise using reusable circuit ``gadgets,'' tensor-network warm starts, and curriculum-based training.

Prior work on differentiable quantum architecture search (DQAS)~\cite{zhang2022dqas} pioneered the use of gradient-based optimization for automated circuit design by representing candidate architectures through a probabilistic model over discrete gate choices. DQAS demonstrated success in unitary decomposition, error mitigation, and QAOA layout discovery by estimating gradients via Monte Carlo sampling from a parameterized distribution. More recently, QuantumDARTS~\cite{wu2023quantumdarts} extended this paradigm using Gumbel-Softmax reparameterization~\cite{jang2017gumbel,maddison2017concrete} to enable end-to-end differentiability without explicit sampling, introducing both macro-search (full circuit) and micro-search (transferable sub-circuit) strategies. While these methods represent significant advances, they treat each circuit placeholder independently and rely on probabilistic gate selection rather than continuous structural interpolation.

In this work, we introduce a framework that bridges the gap between symbolic logic and differentiable programming to create a flexible, gradient-based approach to quantum circuit design (Fig.~\ref{fig:workflow}). 
We are inspired by advances in neuro-symbolic reasoning~\cite{riegel_2020} and differentiable architecture search~\cite{liu_2018}. We treat the existence of each gate in a potential circuit scaffold as a continuous, learnable logical variable $s_i \in [0, 1]$. This formulation allows us to use the power of standard automatic differentiation (i.e., autograd) to optimize the discrete structure of a circuit, and its continuous parameters, simultaneously. Unlike DQAS, which uses Monte Carlo gradient estimation over a probabilistic model, our approach directly interpolates between the identity and gate unitaries via continuous switches, enabling unified structure-parameter optimization through standard backpropagation.

The core of our approach is the ability to train the model not on a single, fixed objective function, but on a set of user-defined, differentiable \textit{logical axioms} that define ``goodness.'' This provides substantial flexibility. By composing different axioms, the same model can function as (1) a compiler with axioms: $\mathcal{T}_{\text{fid}}$, $\mathcal{T}_{\text{simp}}$, which seeks a short gate sequence for a target unitary; (2) a ``VQE Designer'' with axioms: $\mathcal{L}_{\text{energy}}$, $\mathcal{L}_{\text{simp}}$, which learns a compact, parameterized ansatz for a ground state; and (3) a ``robust designer'' with axioms: $\mathcal{L}_{\text{energy}}$, $\mathcal{L}_{\text{rob}}$, $\mathcal{L}_{\text{simp}}$, which performs multi-objective optimization to balance ideal performance, noise resilience, and gate count.

The main benefit of this framework is its ease of implementation and generality. It recasts a discrete, $NP$-hard search problem into a continuous optimization problem solvable with standard deep learning tools. In this paper, we present the theory of this Differentiable Logical Programming (DLP) framework and show that it can discover canonical algorithms, learn compiler optimizations, and design multi-objective VQE circuits from first principles.

The trajectory toward practical quantum advantage is currently obstructed by a significant disconnect between abstract algorithmic theory and the physical constraints of hardware. Babbush et al.~\cite{babbush2025grand} formalize this struggle into a five-stage framework. They identify critical bottlenecks, specifically in the discovery of verifiable algorithms (Stage I) and the optimization required for fault-tolerant compilation (Stage IV). The authors argue that the community faces a ``collective action problem,'' where the scarcity of useful algorithms and the complexities of resource estimation threaten to stall the field's momentum. Crucially, they highlight that simple heuristics for compilation are often insufficient for the emerging early fault-tolerant era.

Our work is a methodological response to these challenges. By reframing discrete circuit synthesis as a continuous, differentiable logic problem, this framework targets the ``scarcity of demonstrated use cases'' discussed by the Google Quantum AI team. Where the Grand Challenge emphasizes the need for algorithms that are robust to noise and physically realizable, our DLP model can learn circuit structures under these constraints, including regimes affected by barren plateaus in variational optimization. This perspective suggests that part of the ``algorithm search'' bottleneck may be addressed with neuro-symbolic architectures that support gradient-based structure learning.

\section{Methodology: A Differentiable Logic for Quantum Circuits}
\label{sec:methodology}

Our framework formalizes circuit discovery as a continuous optimization problem. We first define a scaffold, $S = \{G_1, G_2, \dots, G_N\}$, an ordered list of candidate gate operations that forms a superset of the (unknown) optimal circuit.

\subsection{The Differentiable Circuit Scaffold}

For each candidate gate $G_i \in S$, we associate a learnable real-valued parameter $\lambda_i \in \mathbb{R}$, which we refer to as a ``structural logit.'' The complete set of structural logits forms the learnable parameter vector $\boldsymbol{\lambda} = (\lambda_1, \lambda_2, \dots, \lambda_N) \in \mathbb{R}^N$. Each logit is mapped to a continuous ``gate switch'' $s_i \in [0, 1]$ via the sigmoid function:
\begin{equation}
    s_i = \sigma(\lambda_i) = \frac{1}{1 + e^{-\lambda_i}}.
    \label{eq:sigmoid}
\end{equation}
This switch $s_i$ represents the continuous ``truth value'' of the proposition: ``Gate $G_i$ is active in the circuit.'' The sigmoid mapping ensures $s_i \in [0,1]$ while allowing unbounded gradient flow through $\boldsymbol{\lambda}$.

To make the discrete gate selection differentiable, we must define an effective gate $\tilde{G}_i$ dependent on this continuous switch. We propose two formulations for this interpolation: \textit{Linear Relaxation} and \textit{Geodesic Interpolation}.

\textbf{Formulation 1: Linear Relaxation.}
The simplest approach, used primarily for its computational efficiency, interpolates linearly between the identity and the gate unitary:
\begin{equation}
    \tilde{G}_i^{\text{lin}}(s_i) = (1 - s_i) \mathbb{I} + s_i G_i.
    \label{eq:effective_gate_lin}
\end{equation}
While intermediate states $s_i \in (0, 1)$ are not strictly unitary (potentially dampening the state vector norm), the endpoint solutions ($s_i \in \{0, 1\}$) are guaranteed to be valid quantum gates. We find this method sufficient for shallow depth optimization.

\textbf{Formulation 2: Geodesic Interpolation.}
For deeper circuits where maintaining strict unitarity is critical to prevent numerical instability (vanishing gradients due to norm decay), we define the effective gate via the exponential map, creating a path along the unitary manifold:
\begin{equation}
    \tilde{G}_i^{\text{geo}}(s_i) = G_i^{s_i} = \exp(-i H_i s_i \theta),
    \label{eq:effective_gate_geo}
\end{equation}
where $G_i = e^{-i H_i \theta}$. In this work, we primarily employ Linear Relaxation due to its convexity properties near the boundaries, but note that Eq.~\ref{eq:effective_gate_geo} offers a strictly physical alternative.

The total learned unitary for the entire scaffold, $U_{\text{learned}}$, is the ordered product of all effective gates. Note that matrix multiplication is non-commutative; we define the product order to match the circuit diagram (gates $0 \dots N$ applied left-to-right):
\begin{equation}
    U_{\text{learned}}(\boldsymbol{\lambda}, \boldsymbol{\theta}) = \tilde{G}_N(s_N) \cdot \dots \cdot \tilde{G}_1(s_1) \cdot \tilde{G}_0(s_0),
    \label{eq:full_unitary}
\end{equation}
where $s_i = \sigma(\lambda_i)$ and $\boldsymbol{\theta}$ denotes any continuous rotation angles in parameterized gates.

\subsection{Core Differentiable Axioms}
\label{sec:axioms}

The core of our logical programming approach is that the model is trained to satisfy a set of logical axioms, $\{\mathcal{A}_k\}$, which encode the definition of a ``good'' circuit. Each axiom $\mathcal{A}_k$ is formulated as a differentiable predicate $\mathcal{T}_k \in [0, 1]$, which measures its degree of truth. Here we define the primary axioms used throughout this work.

\paragraph{Correctness (Fidelity).}
Used for compiler/synthesis tasks with a known target unitary $U_{\text{target}}$. The predicate $\mathcal{T}_{\text{fid}}$ measures the normalized squared trace fidelity:
\begin{equation}
    \mathcal{T}_{\text{fid}}(U) = \frac{1}{d^2} \left| \text{Tr}(U_{\text{target}}^\dagger U) \right|^2,
    \label{eq:fidelity_pred}
\end{equation}
where $d = 2^N$ is the matrix dimension. The corresponding loss (contradiction) is:
\begin{equation}
    \mathcal{L}_{\text{fid}} = 1 - \mathcal{T}_{\text{fid}}.
    \label{eq:fidelity_loss}
\end{equation}

\paragraph{Correctness (Energy).}
Used for VQE/QAOA tasks where the goal is to minimize the energy $E = \langle \psi | H | \psi \rangle$ for a state $|\psi\rangle = U|0\rangle$:
\begin{equation}
    \mathcal{L}_{\text{energy}}(\boldsymbol{\lambda}, \boldsymbol{\theta}) = \langle 0 | U(\boldsymbol{\lambda}, \boldsymbol{\theta})^\dagger H \, U(\boldsymbol{\lambda}, \boldsymbol{\theta}) |0 \rangle.
    \label{eq:energy_loss}
\end{equation}

\paragraph{Simplicity (Cost-Weighted).}
The axiom ``The circuit must be simple,'' used for pruning. The predicate applies an exponential penalty on the total cost:
\begin{equation}
    \mathcal{T}_{\text{simp}}(\boldsymbol{s}) = \exp\left(-\alpha \sum_{i=1}^N c_i s_i\right),
    \label{eq:simp_pred}
\end{equation}
where $\alpha$ is a hyperparameter and $c_i$ is the pre-defined cost of gate $G_i$. In practice, we often use the linear approximation for the loss directly, as it provides a constant, stabilizing gradient:
\begin{equation}
    \mathcal{L}_{\text{simp}} = \sum_{i=1}^N c_i s_i.
    \label{eq:simp_loss}
\end{equation}

Additional axioms for entanglement and noise robustness are provided in Appendix~\ref{app:axioms}.

\subsection{Gradient Propagation and Barren Plateaus}
The entire model is optimized using standard gradient descent. The gradient $\nabla \mathcal{L}$ propagates back to the structural logits $\boldsymbol{\lambda}$ (and angle parameters $\boldsymbol{\theta}$) via the chain rule. For a structural logit $\lambda_i$, the gradient is:
\begin{equation}
    \frac{\partial \mathcal{L}}{\partial \lambda_i} = \frac{\partial \mathcal{L}}{\partial s_i} \cdot \frac{\partial s_i}{\partial \lambda_i} = \frac{\partial \mathcal{L}}{\partial s_i} \cdot \sigma(\lambda_i)(1-\sigma(\lambda_i)),
\end{equation}
where the second factor is the derivative of the sigmoid function.

A significant theoretical challenge in variational quantum algorithms is the ``Barren Plateau'' problem, where the variance of the cost function gradients vanishes exponentially with the number of qubits, $\text{Var}(\partial_\theta \mathcal{L}) \propto O(2^{-N})$~\cite{mcclean_2018}. This typically occurs when circuits are initialized as random unitary 2-designs. A ``polluted'' scaffold with random initial parameters effectively acts as such a random circuit.

To mitigate this, we employ a biased initialization strategy. We initialize the structural logits such that the initial switches are biased towards the identity ($\lambda_i \ll 0 \implies s_i \approx 0$). This ensures the optimization trajectory begins in a region of the landscape with non-vanishing gradients (close to the Identity), effectively ``growing'' the circuit complexity only as required by the logical axioms.

\section{Theoretical Analysis of Linear Relaxation}
\label{sec:theory_convergence}

A central component of our Differentiable Logical Programming (DLP) framework is the linear relaxation of the discrete gate choice. We define the effective operation for the $i$-th candidate gate as:
\begin{equation}
    \tilde{G}_i(s_i) = (1-s_i)\mathbb{I} + s_i U_i, \quad s_i \in [0,1].
    \label{eq:linear_relax}
\end{equation}
For intermediate values $s_i \in (0,1)$, the operator $\tilde{G}_i$ is generally non-unitary, i.e., $\tilde{G}_i^\dagger \tilde{G}_i \neq \mathbb{I}$, resulting in a non-physical evolution where the norm of the state vector $|\psi\rangle$ is not preserved. We analyze this formulation as an optimization surrogate, establishing bounds on the induced error and the conditions under which the optimizer is driven toward valid quantum circuits.

\subsection{Norm Deviation Bound}

We first quantify the deviation from unitarity for a single interpolated gate. For any unitary $U_i$ and switch value $s_i \in [0,1]$, the operator $\tilde{G}_i(s_i) = (1-s_i)\mathbb{I} + s_i U_i$ satisfies:
\begin{equation}
    \tilde{G}_i^\dagger \tilde{G}_i = \mathbb{I} + s_i(1-s_i)\left(U_i + U_i^\dagger - 2\mathbb{I}\right).
    \label{eq:norm_deviation}
\end{equation}

\begin{proof}
Expanding directly:
\begin{align}
    \tilde{G}_i^\dagger \tilde{G}_i &= \left[(1-s_i)\mathbb{I} + s_i U_i^\dagger\right]\left[(1-s_i)\mathbb{I} + s_i U_i\right] \nonumber \\
    &= (1-s_i)^2 \mathbb{I} + s_i(1-s_i)(U_i + U_i^\dagger) + s_i^2 \mathbb{I} \nonumber \\
    &= \left[(1-s_i)^2 + s_i^2\right]\mathbb{I} + s_i(1-s_i)(U_i + U_i^\dagger) \nonumber \\
    &= \left[1 - 2s_i(1-s_i)\right]\mathbb{I} + s_i(1-s_i)(U_i + U_i^\dagger) \nonumber \\
    &= \mathbb{I} + s_i(1-s_i)\left(U_i + U_i^\dagger - 2\mathbb{I}\right). \qedhere
\end{align}
\end{proof}

The maximum deviation from unitarity is thus controlled by the factor $s_i(1-s_i) \leq 1/4$, which vanishes at the endpoints $s_i \in \{0,1\}$. In operator norm, this yields:
\begin{equation}
    \left\|\tilde{G}_i^\dagger \tilde{G}_i - \mathbb{I}\right\| \leq s_i(1-s_i) \left\|U_i + U_i^\dagger - 2\mathbb{I}\right\| \leq 4s_i(1-s_i),
    \label{eq:norm_bound}
\end{equation}
where the second inequality uses $\|U_i + U_i^\dagger - 2\mathbb{I}\| \leq 4$ (since the eigenvalues of $U_i$ lie on the unit circle).

For a cascade of $N_S$ gates with switches $\boldsymbol{s} = (s_1, \dots, s_{N_S})$, the total norm deviation of the state $|\psi\rangle = \tilde{G}_{N_S} \cdots \tilde{G}_1 |0\rangle$ can be bounded by:
\begin{equation}
    \left| \langle \psi | \psi \rangle - 1 \right| \leq \prod_{i=1}^{N_S} \left(1 + 4s_i(1-s_i)\right) - 1.
    \label{eq:cascade_bound}
\end{equation}
When all switches are near the binary endpoints (i.e., $|s_i - s_i^*| < \epsilon$ for $s_i^* \in \{0,1\}$), this simplifies to $O(N_S \epsilon)$, confirming that the non-unitarity is well-controlled in the regime where the optimizer operates after initial convergence.

\subsection{Gradient Stability}

Unlike geodesic interpolation ($e^{-iHs\theta}$), whose gradient involves the matrix exponential and can suffer from spectral crowding, linear interpolation provides a constant Jacobian:
\begin{equation}
    \frac{\partial \tilde{G}_i}{\partial s_i} = U_i - \mathbb{I},
    \label{eq:constant_grad}
\end{equation}
independent of $s_i$. This ensures that the gradient signal does not vanish as $s_i \to 0$ (near identity), addressing the ``vanishing gradient'' problem that plagues parameterized unitary circuits near the identity~\cite{mcclean_2018}. The gradient of the full fidelity loss with respect to a structural logit $\lambda_i$ is:
\begin{equation}
    \frac{\partial \mathcal{L}_{\text{fid}}}{\partial \lambda_i} = \underbrace{\frac{\partial \mathcal{L}_{\text{fid}}}{\partial \tilde{G}_i}}_{\text{chain rule}} \cdot \underbrace{(U_i - \mathbb{I})}_{\text{constant}} \cdot \underbrace{s_i(1 - s_i)}_{\text{sigmoid derivative}},
    \label{eq:full_grad}
\end{equation}
where the sigmoid derivative $s_i(1-s_i)$ is the only $s_i$-dependent factor. This means the structural gradient is well-conditioned whenever $s_i$ is not already saturated at 0 or 1, with the saturation being controlled by the logit initialization.

\subsection{Implicit Binarization by the Fidelity Objective}

We now argue that the fidelity objective $\mathcal{L}_{\text{fid}} = 1 - \frac{1}{d^2}|\mathrm{Tr}(U_{\text{target}}^\dagger U_{\text{learned}})|^2$ implicitly drives the switches toward binary values. Consider a single switch $s_i$ with all others fixed. The learned unitary can be decomposed as $U_{\text{learned}} = A \tilde{G}_i(s_i) B$, where $A$ and $B$ are the products of other effective gates. The fidelity becomes:
\begin{equation}
    \mathcal{T}_{\text{fid}}(s_i) = \frac{1}{d^2}\left| (1-s_i)\mathrm{Tr}(M_I) + s_i \mathrm{Tr}(M_U)\right|^2,
    \label{eq:fid_decomp}
\end{equation}
where $M_I = U_{\text{target}}^\dagger A B$ and $M_U = U_{\text{target}}^\dagger A U_i B$. This is a quadratic function in $s_i$, whose maximum over $[0,1]$ is attained at a boundary point whenever $|\mathrm{Tr}(M_I)| \neq |\mathrm{Tr}(M_U)|$ (i.e., whenever including or excluding the gate makes a difference to fidelity). The simplicity penalty $\mathcal{L}_{\text{simp}} = \sum c_i s_i$ adds a linear term that further biases toward $s_i = 0$, reinforcing binarization. Together, these two forces ensure that at convergence, the switches satisfy $s_i \in \{0, 1\}$ up to numerical precision for all gates that meaningfully affect the target fidelity.

\subsection{Summary}
The linear relaxation serves as a valid optimization surrogate because: (i) the norm deviation is bounded by $O(s_i(1-s_i))$ per gate and vanishes at the binary endpoints (Eq.~\ref{eq:norm_bound}); (ii) the constant Jacobian (Eq.~\ref{eq:constant_grad}) provides stable, non-vanishing gradients throughout training; and (iii) the fidelity and simplicity objectives jointly drive the switches toward $\{0,1\}$, where strict unitarity is recovered (Eq.~\ref{eq:fid_decomp}). For deep circuits where intermediate norm decay poses numerical risks, the geodesic interpolation (Eq.~\ref{eq:effective_gate_geo}) provides a strictly unitary alternative.

\subsection{Joint Structure-Parameter Optimization}

The framework is not limited to discrete structural search. Many quantum gates, particularly in VQE and QAOA, are parameterized, e.g., $R_y(\theta)$. We can incorporate these continuous parameters $\boldsymbol{\theta}$ directly into our model.

The scaffold $S$ can contain parameterized gates $G_i(\theta_i)$. The effective gate definition is thus extended:
\begin{equation}
    \tilde{G}_i(s_i, \theta_i) = (1 - s_i)\mathbb{I} + s_i G_i(\theta_i).
    \label{eq:param_gate}
\end{equation}
The total learned unitary $U_{\text{learned}}(\boldsymbol{\lambda}, \boldsymbol{\theta})$ is now a function of both the structural logits $\boldsymbol{\lambda}$ (which determine the switches $\boldsymbol{s}$) and the continuous angles $\boldsymbol{\theta}$. The optimization process will learn both simultaneously. This unified approach is a significant advantage over methods that must alternate between discrete structural updates and continuous parameter optimization.

\subsection{A Differentiable Axiom System via T-Norms}

We can interpret the axiom framework theoretically through the lens of Fuzzy Logic. We treat the problem as finding the Maximum Satisfiability (MaxSAT) of a logical formula. The total loss function $\mathcal{L}_{\text{total}}$ corresponds to the negation of the weighted conjunction of our axioms. Using the \textit{Lukasiewicz T-norm} for conjunction ($\mathcal{T}(x, y) = \max(0, x+y-1)$), the logical contradiction is minimized by minimizing the sum of individual violations:
\begin{equation}
    \mathcal{L}_{\text{total}} = \sum_k w_k \mathcal{L}_k = \sum_k w_k (1 - \mathcal{T}_k).
    \label{eq:total_loss}
\end{equation}
By optimizing $\boldsymbol{\lambda}$ (the structural logits determining gate inclusion) and $\boldsymbol{\theta}$ (the continuous rotation angles) to minimize $\mathcal{L}_{\text{total}}$, the system performs a gradient-based search over the high-dimensional continuous space of circuit architectures, converging on a solution that satisfies the logical proposition of a ``valid circuit.''

\subsection{Complexity, Scalability, and Limitations}
The computational complexity of this framework has two primary components. The forward pass (constructing $U_{\text{learned}}$) involves $N_S$ matrix-matrix multiplications, where $N_S$ is the number of gates in the scaffold. The backward pass (computing gradients) has a similar cost. The dominant computational bottleneck is the $O(d^3) = O(2^{3N})$ complexity of multiplying the $2^N \times 2^N$ unitary matrices, where $N$ is the number of qubits. The cost scales linearly with $N_S$.

This exponential scaling with qubit count $N$ is the fundamental limitation of all full-state quantum simulation methods. It makes the ``flat'' (non-hierarchical) application of this framework intractable for $N \gtrsim 16$ qubits. This limitation is the primary motivation for the Hierarchical Synthesis (HS) framework, which we detail in Section~\ref{sec:hs}.

\subsection{Methodology for Complex Problems: Curricula}
\label{sec:curricula}
A known challenge in high-dimensional optimization is the presence of poor local minima. These are distinct from barren plateaus (addressed in Section~2.3): while barren plateaus cause gradient magnitudes to vanish exponentially, local minima trap the optimizer in suboptimal solutions where gradients exist but point toward inferior configurations. We employ two curriculum learning strategies to guide the optimizer past such local minima toward a global solution.

\paragraph{Annealing Curriculum (for VQE):} For complex VQE problems, optimizing $H$ directly can trap the model in a simple, high-energy state. We instead define an annealing path $H(t) = (1-t)H_{\text{easy}} + tH_{\text{hard}}$, where $H_{\text{easy}}$ is a simple, solvable Hamiltonian (e.g., $t=0$, just the external field) and $H_{\text{hard}}$ is the full target Hamiltonian. By slowly increasing $t$ from 0 to 1 during training, we guide the optimizer from a simple solution to the complex one, successfully avoiding local minima.
\paragraph{Two-Phase Curriculum (for QAOA):} For QAOA depth discovery, we separate ``discovery'' from ``pruning.'' It consists of two phases: 
In Phase 1 (Discovery), we set $w_{\text{simp}} = 0$ and optimize only for energy ($\mathcal{L} = \mathcal{L}_{\text{energy}}$). This allows the model to ``turn on'' all potentially useful layers and find the best possible angles, regardless of cost.
In Phase 2 (Pruning), we turn on the simplicity weight ($w_{\text{simp}} > 0$). The optimizer now prunes any layers that do not contribute significantly to the final energy, discovering the minimal required depth $p$.

\subsection{Methodology for Scalability: Hierarchical Synthesis}
\label{sec:hs}
Simulating the full unitary $U_{\text{learned}}$ (Eq.~\ref{eq:full_unitary}) requires matrices of size $2^N \times 2^N$, which is intractable for $N \gtrsim 16$ qubits. To overcome this, we adopt a Hierarchical Synthesis (HS) framework.

The HS framework breaks a large $N$-qubit problem into a series of small, tractable $n$-qubit sub-problems.
\begin{enumerate}
    \item \textbf{Level 0 (Motif Discovery):} We first use the DLP framework to discover and optimize a small, fundamental $n_0$-qubit motif, $U_{M_0}$ (e.g., a 2-qubit Bell state circuit).
    \item \textbf{Level 1 (Composition):} We ``freeze'' the discovered $U_{M_0}$ and treat it as a new, non-learnable gate. We then promote it to an $n_1$-qubit scaffold $S_1$ (e.g., $U_{M_0}(q_0, q_1) \otimes \mathbb{I}_{q_2}$). The DLP optimizer then solves this $n_1$-qubit problem, discovering $U_{M_1}$.
    \item \textbf{Level $k$ (Iteration):} This process is repeated. The circuit $U_{M_k}$ is composed of motifs $U_{M_{k-1}}$ and other gates.
\end{enumerate}
This hierarchical approach ensures that the computational complexity at each optimization step $k$ scales only with the size of the new sub-problem, $O(2^{n_k})$, rather than the total number of qubits, $O(2^N)$.

\subsection{Comparison to Prior Art \& Hardware Readiness}
This DLP framework differs significantly from non-differentiable search methods, such as genetic algorithms~\cite{evolving_circuits_2002} or symbolic SAT-solvers~\cite{amy_2013_sat,meuli_2019_sat}. Those methods must rely on heuristic search strategies or sampling from a discrete space, which can be highly inefficient. By contrast, our framework maps the discrete search space onto a continuous manifold, allowing for the use of highly efficient, gradient-based optimizers that can follow the path of steepest descent.

Compared to prior differentiable approaches such as DQAS~\cite{zhang2022dqas} and QuantumDARTS~\cite{wu2023quantumdarts}, our method offers several distinctions. DQAS estimates gradients through Monte Carlo sampling from a probabilistic model over discrete architectures, which can be sample-inefficient. QuantumDARTS improves upon this with Gumbel-Softmax reparameterization~\cite{jang2017gumbel,maddison2017concrete} but still frames the problem as probabilistic gate selection at each placeholder. Our approach, by contrast, uses continuous switches $s_i$ that directly interpolate the gate unitary with the identity (Eq.~\ref{eq:effective_gate_lin}), enabling true end-to-end differentiability through standard backpropagation. Furthermore, our logical axiom framework (Section~\ref{sec:axioms}) provides a principled way to compose multiple objectives (fidelity, simplicity, robustness) that extends beyond the single-objective formulations typical of prior QAS methods.

The circuits discovered by this method are not guaranteed to be ``hardware-native.'' For example, a discovered circuit may contain a $CNOT(0,3)$ gate, which is not directly executable on hardware with only linear connectivity. A final post-compilation and routing pass would still be required. However, the $\mathcal{L}_{\text{simp}}$ axiom (Eq.~\ref{eq:simp_loss}) can be made ``hardware-aware'' by assigning costs $c_i$ based on hardware constraints. For example, a native $CNOT(0,1)$ could be assigned $c_i=10$, while a non-native $CNOT(0,3)$ (which requires multiple SWAPs) could be assigned $c_i=70$. This would intrinsically bias the optimizer to discover circuits that are more amenable to a specific target topology.

\section{Experiments and Results}
\label{sec:experiments}

\subsection{Experiment 1: Trotter-Step Optimization}
We demonstrate the framework's capability to perform compiler optimization by discovering a 2nd-order Trotter-Suzuki decomposition from a ``polluted'' scaffold containing redundant and suboptimal gates.

\subsubsection{Implementation Details} 
We define a target unitary $U_{\text{target}} = e^{-i H_{\text{Heis}} t}$ for a 4-qubit 1D Heisenberg chain with $t=0.1$. The goal is to approximate this evolution using discrete gates. We construct a scaffold $S$ containing a mix of valid and invalid operators:
\begin{enumerate}
    \item \textbf{Correct 2nd-Order Gates:} The optimal symmetric sequence $e^{-i H_{\text{odd}} t/2}$, $e^{-i H_{\text{even}} t}$, and $e^{-i H_{\text{odd}} t/2}$.
    \item \textbf{Distractor Gates:} A 1st-order approximation term $e^{-i H_{\text{odd}} t}$, a single-term operator $e^{-i H_{01} t}$, and a generic CNOT gate.
\end{enumerate}

The DLP model is initialized with all gates ``ON'' (logits $\lambda_i = 2.0$, corresponding to $s_i \approx 0.88$) to simulate a pruning task. We train for 4000 epochs using AdamW optimizer with a learning rate of 0.01. The loss function balances correctness (fidelity) and simplicity (gate count):
\begin{equation}
\mathcal{L} = w_{\text{fid}}\mathcal{L}_{\text{fid}} + w_{\text{simp}}\mathcal{L}_{\text{simp}}
\label{eq:trotterloss}
\end{equation}
where $w_{\text{fid}}=5.0$ and $w_{\text{simp}}=0.3$, and the loss terms are defined in Eqs.~\ref{eq:fidelity_loss} and \ref{eq:simp_loss}. This weighting prioritizes finding a high-fidelity solution while exerting constant pressure to remove unnecessary gates.

\begin{figure}[htbp]
    \centering
    \includegraphics[width=0.48\textwidth]{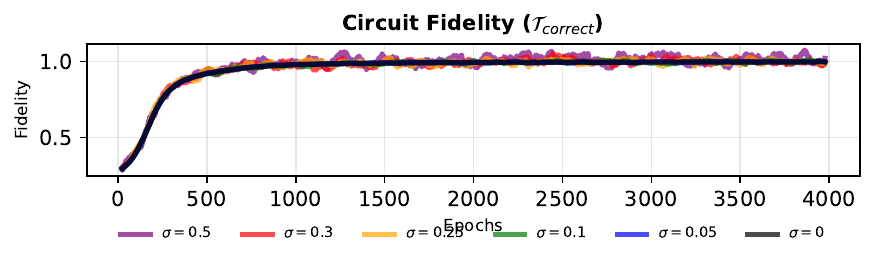}
    \caption{\textbf{Circuit fidelity over training.} Circuit fidelity ($\mathcal{T}_{\text{fid}}$) over training epochs for varying noise levels $\sigma$. The system consistently converges to high fidelity even under significant noise ($\sigma=0.5$).}
    \label{fig:trotter_fidelity}
\end{figure}

\begin{figure}[htbp]
    \centering
    \includegraphics[width=0.48\textwidth]{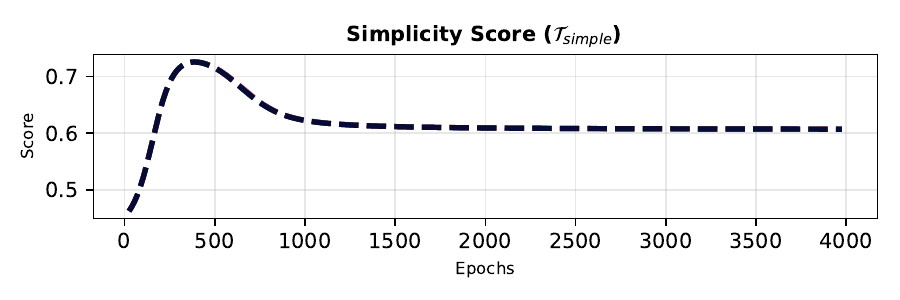}
    \caption{\textbf{Simplicity score over training.} Simplicity score ($\mathcal{T}_{\text{simp}}$) over training epochs. The convergence to a specific score indicates the pruning of redundant gates across all noise levels.}
    \label{fig:trotter_simplicity}
\end{figure}

\subsubsection{Evaluation Criteria}
\paragraph{Structural Accuracy:} The model must identify and keep only the three gates corresponding to the 2nd-order decomposition ($U_{\text{odd}}(t/2), U_{\text{even}}(t), U_{\text{odd}}(t/2)$) while suppressing all distractors (i.e., $s_i < 0.01$).
\paragraph{Fidelity:} The fidelity of the discovered circuit must match the theoretical limit of the 2nd-order approximation ($F > 0.999$).

\paragraph{Robustness:} The discovery process must converge to the correct topology even in the presence of noise.

\subsubsection{Evaluation}
To verify the method's resilience to control and readout errors, we introduced Gaussian noise to the unitary evaluations during training. We performed independent trials with noise levels $\sigma \in \{0, 0.05, 0.1, 0.25, 0.3, 0.5\}$. In every trial, including the highest noise setting ($\sigma=0.5$), the DLP model converged to the optimal 3-gate structure (Figs.~\ref{fig:trotter_fidelity} and \ref{fig:trotter_simplicity}). The discrete nature of the gate switches ($s_i \in \{0, 1\}$) acts as a noise filter: once a gate is decisively pruned or selected, small fluctuations in the loss landscape due to noise are insufficient to reverse the decision. This ``locking-in'' effect, combined with the gradient averaging inherent in stochastic optimization, allows the framework to extract the correct symbolic structure even from a noisy signal.

\begin{figure}[htbp]
    \centering
    \includegraphics[width=\linewidth]{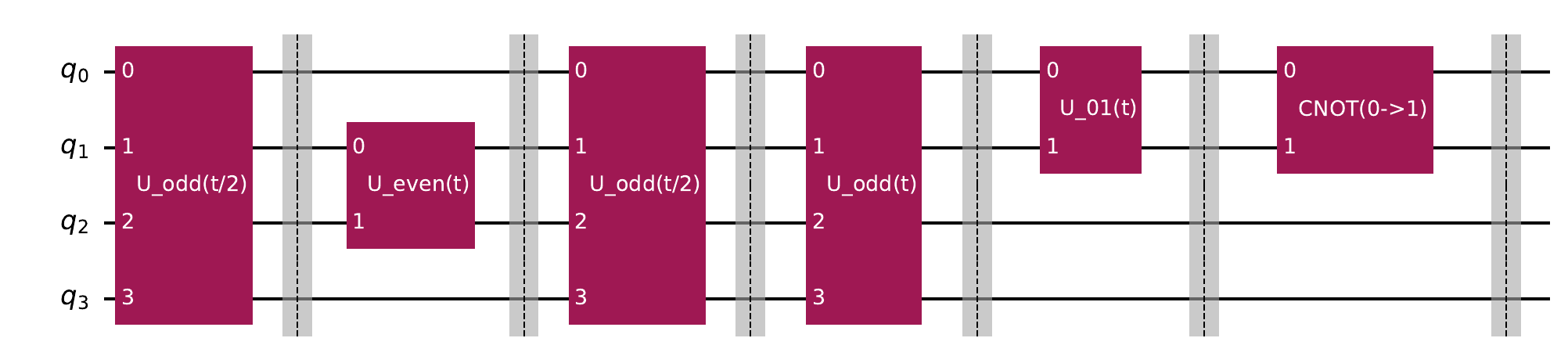}
    \caption{\textbf{Initial scaffold configuration.} Initial scaffold input with all gates activated. This represents the starting point for the pruning task.}
    \label{fig:scaffold_input}
\end{figure}

\begin{figure}[htbp]
    \centering
    \includegraphics[width=\linewidth]{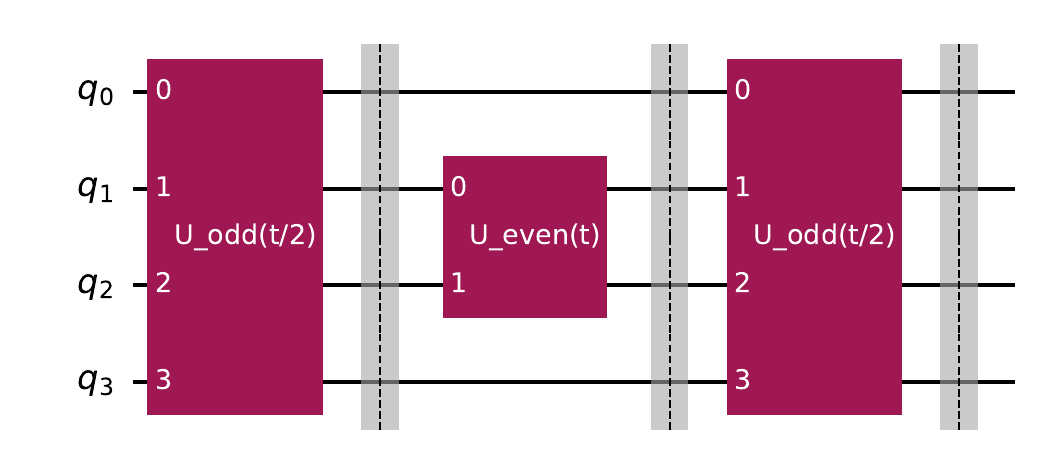}
    \caption{\textbf{Discovered circuit topology.} Discovered circuit topology for noise level $\sigma=0.5$. The framework correctly identifies the 2nd-order Trotter decomposition.}
    \label{fig:scaffold_result}
\end{figure}

Figure~\ref{fig:scaffold_input} illustrates the initial state of the scaffold before optimization, where all candidate gates are active. Figure~\ref{fig:scaffold_result} visualizes the final discovered circuit topology for the highest noise level $\sigma=0.5$, demonstrating the consistency of the structural solution.

\subsubsection{Cost-Aware Non-Trivial Pruning}
We next tested whether the model can make cost-aware pruning decisions that go beyond simple rule-based pattern matching.
\paragraph{Setup:}
The scaffold is a 14-gate, 5-qubit circuit with two distinct redundancies: (a) a visually obvious $H$-$H = I$ identity (cost = 1.0 + 1.0 = 2.0) and (b) a non-trivial $CNOT$-$CNOT = I$ identity (cost = 10.0 + 10.0 = 20.0). The target is $U_{\text{target}}$ of the full 14-gate circuit.

\paragraph{Objective:}
The axioms are $\mathcal{L} = w_{\text{fid}}\mathcal{L}_{\text{fid}} + w_{\text{simp}}\mathcal{L}_{\text{simp}}(\boldsymbol{s}, \boldsymbol{c})$. This is analogous to Eq.~\ref{eq:trotterloss}, but with an explicit dependence on $(\boldsymbol{s}, \boldsymbol{c})$ in the simplicity term, and explicitly tests whether the optimizer prioritizes removing high-cost redundancies.

\paragraph{Result}
The model did not prune the obvious, low-cost $H$-$H$ identity. Instead, it converged on pruning the two CNOT gates, correctly identifying the optimization path that yielded a 10$\times$ greater reduction in the simplicity loss $\mathcal{L}_{\text{simp}}$.

\subsection{Experiment 2: Minimal Circuit Discovery}

To validate the combinatorial search capabilities of the DLP framework, we conducted an experiment targeting the discovery of the 4-qubit Quantum Fourier Transform (QFT) circuit from a polluted scaffold. The primary motivation was to demonstrate that the model could autonomously extract a canonical algorithmic structure from a high-dimensional, noisy search space.

The central challenge lay in the complexity of the input scaffold (Fig.~\ref{fig:qft_results} (a)). We constructed a heavily polluted scaffold containing 21 potential gate operations, corresponding to a large discrete configuration space (which our method explores via continuous relaxation and gradient descent, not brute-force enumeration). This input included the 12 gates required for the optimal QFT, but they were obscured by 9 ``polluter'' gates. These distractors were engineered to be deceptive; they included redundant identity sequences (specifically $H$-$H$ and $CNOT$-$CNOT$ pairs) and other incorrect ``junk'' gates, presenting the optimizer with local minima that a naive pruning method might fail to escape.

The goal was to match the target $16 \times 16$ QFT unitary. The optimization was driven by the composite loss function $\mathcal{L} = w_{\text{fid}}\mathcal{L}_{\text{fid}} + w_{\text{simp}}\mathcal{L}_{\text{simp}}$. This formulation necessitates a delicate balance: the system must maximize logical correctness ($\mathcal{T}_{\text{fid}}$) while simultaneously responding to the pressure to minimize circuit complexity ($\mathcal{L}_{\text{simp}}$).

\subsubsection{Evaluation} 
\begin{figure}[t]
    \centering
    \begin{tabular}{c}
    {\tiny (a) The input scaffold $S$ with 21 gates, including hidden identity pairs.}\\\includegraphics[width=1.0\linewidth]{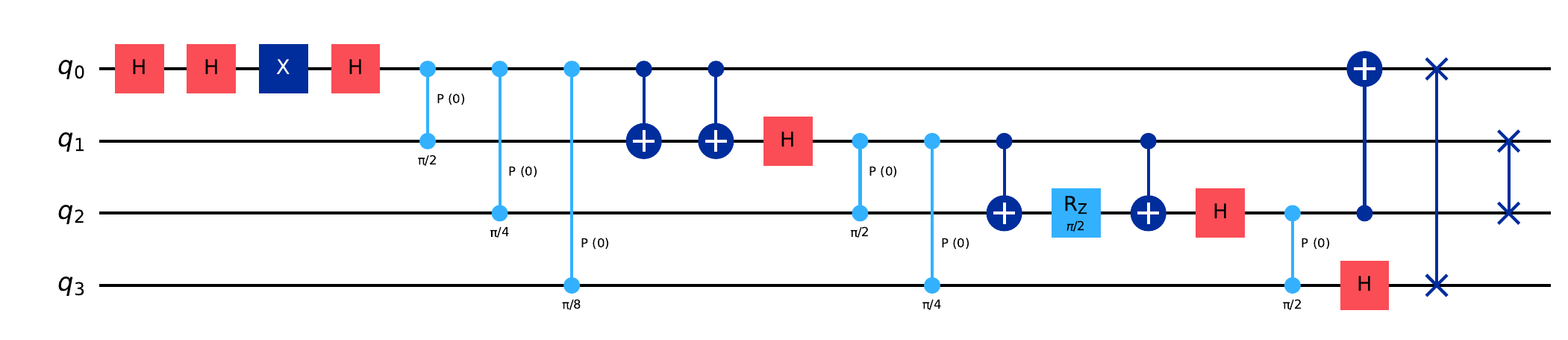}\\
    {\tiny (b) The final discovered circuit, recovering the optimal 12-gate QFT.}\\\includegraphics[width=1.0\linewidth]{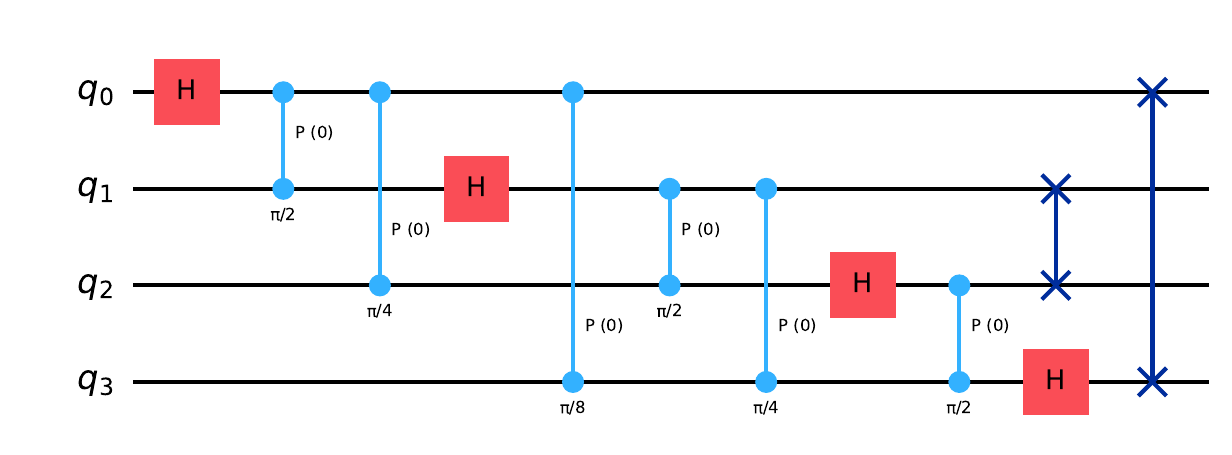}\\
    {\tiny (c) Training dynamics showing convergence of fidelity and cost.}\\\includegraphics[width=1.0\linewidth]{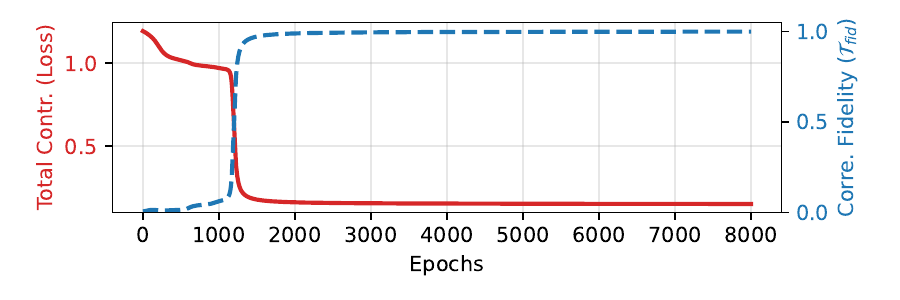}
    \end{tabular}
    \caption{\textbf{Circuit Discovery Results.} The figure illustrates the pruning process for the 4-qubit QFT. \textbf{(a)} The initial state is a ``polluted'' scaffold where valid gates are mixed with distractors (e.g., $H$-$H$, $CNOT$-$CNOT$). \textbf{(b)} The final circuit topology after optimization, where the model has filtered out the noise to reveal the canonical structure. \textbf{(c)} Training curves indicating that the system maximizes fidelity ($\mathcal{T}_{\text{fid}}$) while simultaneously minimizing the simplicity loss ($\mathcal{L}_{\text{simp}}$), optimizing over the discrete configuration space via continuous relaxation.}
    \label{fig:qft_results}
\end{figure}

The results, visualized in Fig.~\ref{fig:qft_results} (b,c), demonstrate a decisive convergence to the optimal solution together with the final circuit. The learned structural switches correctly differentiated between essential and non-essential operations, converging to $s_i \approx 1$ for all 12 correct QFT gates and $s_i \approx 0$ for all 9 polluter gates. By optimizing over this discrete configuration space via continuous relaxation, the framework proved its capacity for non-trivial combinatorial optimization.

\subsection{Experiment 3: Noise Resilience Benchmark Against QuantumDARTS}

We conducted a comprehensive benchmark to validate the framework's robustness to measurement shot noise, comparing DLP against QuantumDARTS on the quantum simulation of the stretched Lithium Hydride (LiH) molecule. Shot noise is a fundamental limitation in NISQ devices arising from the finite number of measurement samples available per circuit evaluation. Our objective was to quantify the noise resilience of both architecture search methods under realistic experimental conditions.
\subsubsection{Problem and Formulation}
We modeled the LiH molecule at a stretched bond distance of $d=\SI{2.5}{\angstrom}$, mapped to 4 qubits using an active space reduction (occupied core: Li 1s; active valence: orbitals 1-2). This geometry exhibits strong static correlation (correlation gap: \SI{2.67}{\milli\hartree}), making it a challenging test case where Hartree-Fock reference is insufficient.

\paragraph{Baseline Scaffold:} Hardware-efficient ansatz (HEA) with 2 layers of $R_Y$, $R_Z$ rotations and linear CNOT entanglers, initialized from the Hartree-Fock state $|1100\rangle$, yielding a search space of $\sim$24 structural degrees of freedom.
\paragraph{Shot Noise Implementation:} For each training iteration, we injected Gaussian noise with variance $\sigma^2 = \text{Var}[H]/n_{\text{shots}}$ to the energy gradient, where $\text{Var}[H] = \langle \psi | H^2 | \psi \rangle - \langle \psi | H | \psi \rangle^2$ is the observable variance. This models the Central Limit Theorem behavior of finite-shot quantum measurements.
\paragraph{Curriculum Learning:} Soft-Pruning Curriculum with 150-epoch warmup ($w_{\text{simp}}=0$) followed by gradual sparsity ramp-up ($w_{\text{simp}} \to 0.002$) over 300 total epochs.
\paragraph{Noise Levels:} We tested six shot regimes: Exact (analytical), 10,000, 1,000, 500, 100, and 50 shots, spanning high-fidelity to severely noisy conditions.

\subsubsection{Results and Discussion}
Table~\ref{tab:shot_noise_results} presents the final ground state errors and circuit sizes discovered under each noise condition.

\begin{table}[htbp]
\centering
\caption{\textbf{Noise Resilience Comparison: DLP vs QuantumDARTS on LiH (\SI{2.5}{\angstrom}).}}
\label{tab:shot_noise_results}
\resizebox{1.0\linewidth}{!}{
\begin{tabular}{l S[table-format=-1.5] S[table-format=2.2] S[table-format=2.0]
                  S[table-format=-1.5] S[table-format=2.2] S[table-format=2.0]}
\toprule
& \multicolumn{3}{c}{\textbf{Ours}} & \multicolumn{3}{c}{\textbf{QuantumDARTS~\cite{wu2023quantumdarts}}} \\
\cmidrule(lr){2-4} \cmidrule(lr){5-7}
\textbf{Shots} & {\textbf{Energy}} & {\textbf{Error}} & {\textbf{Gates}} & {\textbf{Energy}} & {\textbf{Error}} & {\textbf{Gates}} \\
& {\textbf{(Ha)}} & {\textbf{(mHa)}} & & {\textbf{(Ha)}} & {\textbf{(mHa)}} & \\
\midrule
Exact    & -7.77052 &  3.02 &  9 & -7.73624 & 37.30 &  8 \\
10,000   & -7.77064 &  2.91 & 10 & -7.77039 &  3.15 &  9 \\
1,000    & -7.77005 &  3.49 & 11 & -7.77048 &  3.07 & 11 \\
500      & -7.77014 &  3.40 &  9 & -7.70801 & 65.53 &  8 \\
100      & -7.77061 &  2.94 & 11 & -7.76922 &  4.32 & 10 \\
50       & -7.77065 &  2.90 & 11 & -7.77063 &  2.91 & 14 \\
\bottomrule
\end{tabular}}
\end{table}

Our approach exhibits $98.8\times$ lower variance in final energies compared to our implementation of QuantumDARTS across shot noise levels (std: $2.44 \times 10^{-4}$ vs $2.41 \times 10^{-2}$~Ha). This demonstrates that DLP's sigmoid-based continuous switches provide more stable gradients under noisy conditions than QuantumDARTS's Gumbel-Softmax sampling.

The lower variance of our method maintains chemical accuracy ($< \SI{5}{\milli\hartree}$ error) across all noise regimes, with minimal variation in final energy (\SIrange{2.90}{3.49}{\milli\hartree}). 
At the same time, discovered circuit size remained compact (between 9 and 11 gates), indicating stable architectural decisions even under severe noise.

\subsubsection{Interpretation: Smooth vs Sharp Decision Boundaries}

The noise resilience of our approach can be attributed to the difference in how the two methods interpolate between gate inclusion and exclusion:

Our approach uses the sigmoid switch function $s_i = \sigma(\lambda_i)$, which provides a smooth, convex mapping from logits to gate probabilities. The derivative $\partial s_i / \partial \lambda_i = s_i(1-s_i)$ is continuous and bounded, providing stable gradients even when the energy signal is noisy. The sigmoid naturally damps extreme gradient updates, acting as an implicit regularizer.

On the other hand, QuantumDARTS uses reparameterization $w_i = \text{softmax}((\alpha_i + g_i)/\tau)$ (Gumbel-Softmax), where $g_i$ are Gumbel noise samples, introduces stochastic discrete decisions that are sensitive to gradient noise. At low temperatures ($\tau \to 0$), the argmax behavior creates sharp decision boundaries where small gradient perturbations can flip architectural choices, leading to the observed instability.

\subsubsection{Hardware-Aware Optimization Validation}

As a secondary validation, we tested the framework's ability to avoid non-native gates on a simulated 3-qubit linear topology. The optimizer learned to suppress a non-native CNOT(0,2) gate (cost: 101.0, requiring 2 SWAPs) in favor of native CNOT(0,1) and CNOT(1,2) connections (cost: 1.0 each), with the non-native gate's activation probability converging to $s_{0,2} = 0.089 < 0.1$, demonstrating hardware-aware architectural search without explicit rule-based constraints.

\subsection{Experiment 4: Scalable Topology Discovery on the Frustrated Heisenberg Chain}

We applied the Hierarchical Synthesis (HS) framework to address the challenge of \textbf{geometric frustration} in quantum many-body systems. While standard ansatzes often rely on nearest-neighbor connectivity, frustrated systems, such as the 1D Heisenberg chain with competing nearest-neighbor ($J_1$) and next-nearest-neighbor ($J_2$) interactions, require non-local topology to accurately capture the ground state.

\subsubsection{Problem and Formulation}
The objective was to find the ground state of the Hamiltonian:
\begin{equation}
    H = J_1 \sum_i \vec{S}_i \cdot \vec{S}_{i+1} + J_2 \sum_i \vec{S}_i \cdot \vec{S}_{i+2}
\end{equation}
A linear chain topology is insufficient for this system due to the competing interaction terms. Furthermore, optimizing a circuit for a macroscopic system size (e.g., $N=20$) is computationally intractable due to the exponential scaling of the state vector.

To overcome this, we formulated the problem using the HS ``Divide-and-Conquer'' strategy:
\begin{itemize}
    \item \textbf{Stage I (Motif Discovery):} We defined a minimal 3-qubit ``motif'' scaffold ($q_0, q_1, q_2$). The scaffold included parameterizable rotations ($R_Y, R_Z$) and a learnable set of entangling gates corresponding to all possible pairwise interactions: nearest-neighbor ($q_0-q_1, q_1-q_2$) and the non-local next-nearest-neighbor ($q_0-q_2$).
    \item \textbf{Stage II (Hierarchical Compilation):} The discovered motif was ``frozen'' and procedurally tiled to construct a linear array for $N=20$ qubits, bypassing the barren plateau problem associated with global optimization.
\end{itemize}

\subsubsection{Discovery Results}
\begin{figure}[t]
    \centering
    \includegraphics[width=0.95\linewidth]{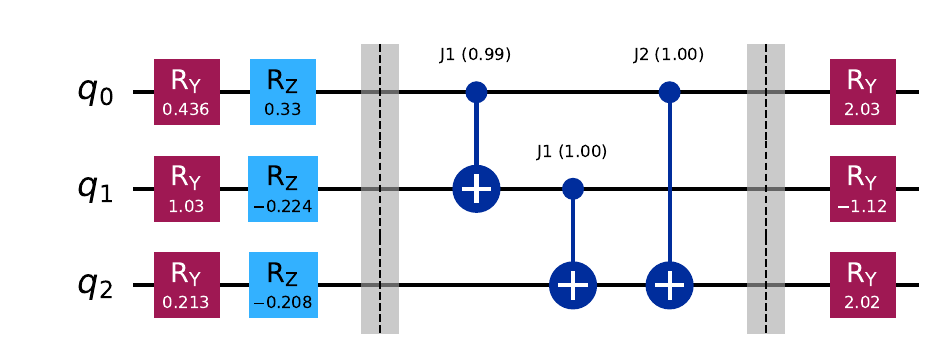}
    \caption{\textbf{Discovered motif for the $J_1$-$J_2$ Heisenberg model.} The DLP framework correctly identified the need for a triangular topology, activating the non-local $J_2$ gate (connection between $q_0$ and $q_2$) alongside standard nearest-neighbor interactions to resolve geometric frustration.}
    \label{fig:j1j2_motif}
\end{figure}

In Stage I, the DLP optimizer was tasked with finding the minimal local structure required to resolve the frustration. Fig.~\ref{fig:j1j2_motif} illustrates the final discovered motif. The optimization produced two key results:

\begin{itemize}
    \item \textbf{Topological Search:} The model autonomously identified the necessity of the non-local interaction. As shown in the circuit diagram, the optimizer converged to a topology where the ``skip'' connection ($q_0$ connected to $q_2$) was fully activated ($s \approx 1.0$), alongside the nearest-neighbor connections. This effectively created a triangular connectivity graph required to resolve the geometric frustration.
    \item \textbf{Parameter Optimization:} The system simultaneously optimized the rotation angles, converging to an energy of $E \approx -3.32$ for the local block.
\end{itemize}

\subsubsection{Scalability and Comparison}
In Stage II, the discovered 3-qubit motif was compiled into a full ansatz for a 20-qubit system. This process generated a physics-informed circuit with a total depth of 216 gates. By leveraging the HS framework, we achieved a result that would have been computationally inaccessible via direct global optimization ($2^{20}$ Hilbert space), demonstrating the framework's capacity to scale physically motivated circuit structures to macroscopic regimes.

\subsection{Experiment 5: Hardware-Aware Optimization of Geometrically Frustrated Systems}

\paragraph{Motivation}
A critical bottleneck in the NISQ era is the disconnect between abstract algorithmic design and physical hardware constraints. While theoretical ansatzes often assume all-to-all connectivity, physical devices typically possess restricted coupling maps (e.g., linear or grid topologies). Mapping a non-native circuit to such hardware requires the insertion of SWAP gates, which increases circuit depth and accumulates coherent errors. This challenge is particularly acute in geometrically frustrated systems, such as the $J_1$-$J_2$ Heisenberg model, where competing interactions often require non-local connectivity to resolve the ground state.

\paragraph{Methodology}
To address these challenges, we applied the framework to the 3-qubit $J_1$-$J_2$ Heisenberg model with parameters $J_1=1.0$ and $J_2=0.5$. We imposed a strict linear device topology ($0-1-2$), rendering the next-nearest-neighbor interaction between qubits 0 and 2 non-native. The optimization objective was defined as a composite loss function:
\begin{equation}
    \mathcal{L} = w_{\text{energy}}\mathcal{L}_{\text{energy}} + w_{\text{hw}}\sum_{i} c_i s_i
\end{equation}
where $c_i$ represents the hardware cost. We assigned a high penalty ($c_{\text{non-native}}=100.0$) to CNOT gates acting on unconnected qubit pairs $(0,2)$, while native gates were assigned a nominal cost ($c_{\text{native}}=1.0$). We compared the hardware-aware DLP against a standard DLP model optimizing purely for energy without topological constraints.

\begin{figure}[ht!]
    \centering
    \includegraphics[width=1.0\linewidth]{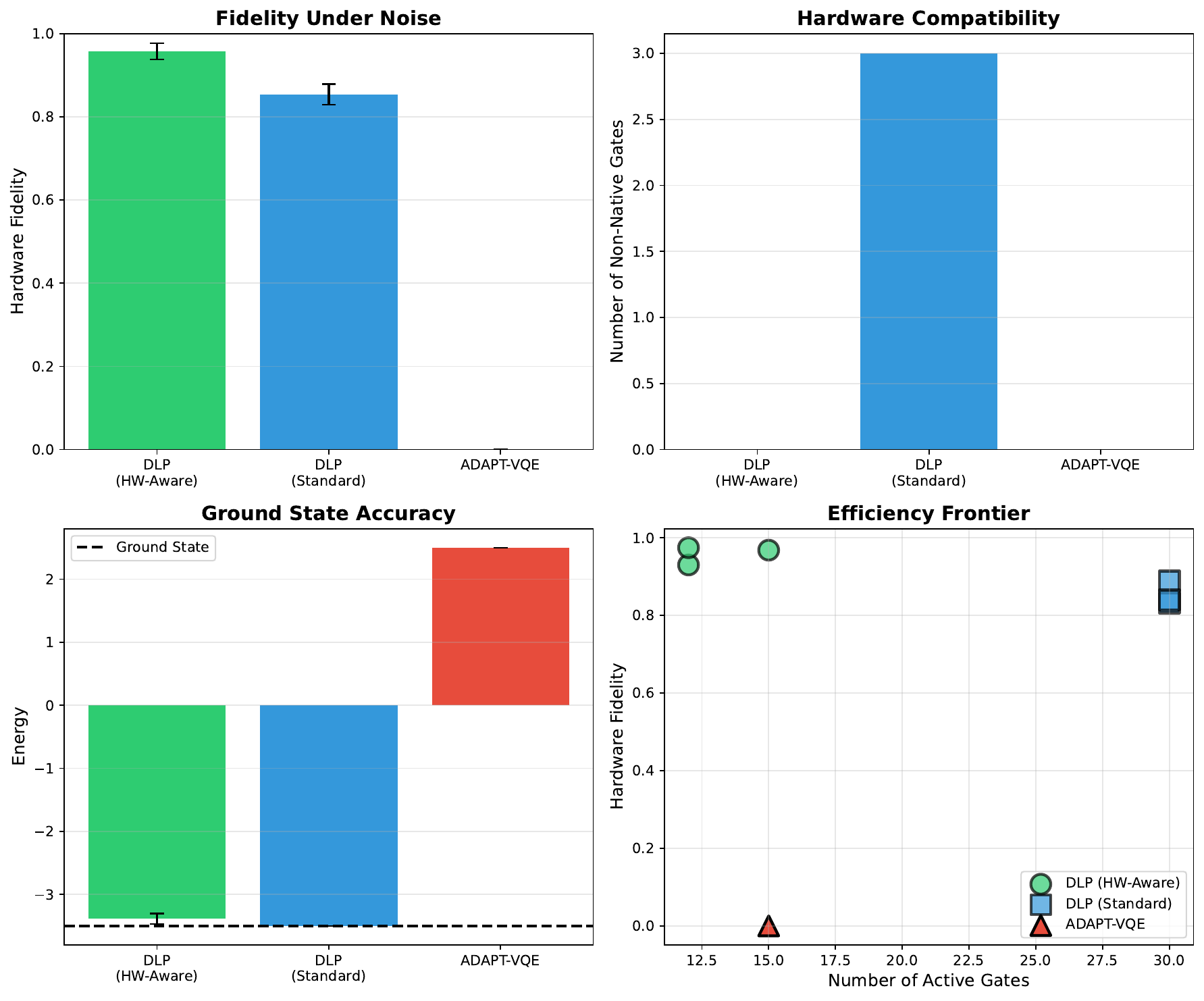}
    \caption{\textbf{Fidelity vs.\ Complexity.} The hardware-aware DLP (green) sacrifices a small amount of theoretical energy to drastically reduce circuit depth by avoiding non-native gates. This results in a $\approx 12\%$ improvement in realized fidelity compared to the standard DLP (blue), which suffers from SWAP-induced errors.}
    \label{fig:fidelity_comparison}
\end{figure}

\paragraph{Results}
The results, summarized in Table~\ref{tab:comparison_results}, demonstrate the efficacy of the hardware-aware approach. Both DLP variants successfully navigated the energy landscape. The standard DLP achieved near-perfect theoretical convergence (99.9--100\% ground state overlap). However, it relied heavily on non-native gates (average of 3.0 per circuit), necessitating significant compilation overhead. As shown in Figure~\ref{fig:fidelity_comparison}, this resulted in a post-compilation depth of 37 and a reduced hardware fidelity of $0.853 \pm 0.025$ due to noise accumulation.

The hardware-aware DLP discovered a topology that avoided the non-native connection entirely (0.0 non-native gates). Although this trade-off resulted in a marginally higher theoretical energy (96.7\% of ground state), the resulting circuits were significantly shallower (Depth 17). This structural efficiency translated to a superior hardware fidelity of $0.957 \pm 0.020$. A t-test confirms this improvement is statistically significant ($t=4.618$, $p=0.0099 < 0.05$). These findings suggest that incorporating hardware constraints directly into the differentiable logic allows the optimizer to find ``hardware-native'' solutions that outperform theoretically optimal but compilation-heavy circuits.

\begin{table}[h]
    \centering
    \caption{\textbf{Benchmarking Hardware-Aware Optimization.} Comparison of training energy (theoretical), compilation overhead, and final hardware fidelity under a noise model.}
    \label{tab:comparison_results}
    \resizebox{1.0\linewidth}{!}{
    \begin{tabular}{lccc}
        \toprule
        \textbf{Method} & \textbf{Ground State Overlap} & \textbf{Compiled Depth} & \textbf{Hardware Fidelity} \\
        \midrule
        Standard DLP & $\mathbf{99.9\text{--}100\%}$ & 37 & $0.853 \pm 0.025$ \\
        \textbf{HW-Aware DLP} & $96.7\%$ & \textbf{17} & $\mathbf{0.957 \pm 0.020}$ \\
        \bottomrule
    \end{tabular}}
\end{table}

\subsection{Experiment 6: Adaptive Routing on IBM Quantum Hardware}
\label{sec:ibm_experiment}

To validate the practical applicability of the DLP framework beyond simulation, we conducted adaptive routing experiments on IBM Quantum's \texttt{ibm\_fez} backend, a 156-qubit Heron~R2 superconducting quantum processor. These experiments demonstrate that the framework can  detect and respond to hardware degradation using only measurement-driven gradient updates, without requiring explicit failure detection protocols, manual intervention, or hardwired path preferences.

\subsubsection{Experimental Setup}

The task is 3-qubit GHZ state preparation ($|\psi\rangle = \frac{1}{\sqrt{2}}(|000\rangle + |111\rangle)$) with two redundant topological paths through the device: Path~A ($0 \to 1 \to 2$, using $\textsc{cx}(0{,}1) \to \textsc{cx}(1{,}2)$) and Path~B ($0 \to 2 \to 1$, using $\textsc{cx}(0{,}2) \to \textsc{cx}(2{,}1)$). The ``scaffold'' is these two mutually exclusive paths.

For mutually exclusive path selection, we employ a \textit{Softmax Router}: a natural extension of the sigmoid gate switches to the multi-choice setting. The structural logits $\boldsymbol{\lambda} = (\lambda_A, \lambda_B) \in \mathbb{R}^2$ are mapped to path probabilities via softmax:
\begin{equation}
    p_i = \frac{e^{\lambda_i}}{\sum_j e^{\lambda_j}}, \quad p_A + p_B = 1.
    \label{eq:softmax_router}
\end{equation}
The fidelity axiom loss is the expected infidelity under the soft path selection:
\begin{equation}
    \mathcal{L} = 1 - \sum_i p_i \cdot F_i,
    \label{eq:routing_loss}
\end{equation}
where $F_i$ is the measured GHZ fidelity on path~$i$. The gradient $\partial \mathcal{L}/\partial \lambda_i = p_i(\bar{F} - F_i)$ naturally pushes probability toward whichever path has higher fidelity. Both paths are measured every cycle (512 shots each), and the Adam optimizer ($\eta=0.5$) performs 10 gradient updates per cycle. The logits are clamped to $[-2, 2]$ (crossing) or $[-3, 3]$ (catastrophic) to prevent saturation while maintaining recovery capability. All logits are initialized to zero and the router has no prior preference and must discover the better path from measurements alone.

\subsubsection{Experiment 6a: Gradual Noise Crossing}
\label{sec:crossing}

\paragraph{Methodology.}
To simulate calibration drift, we inject controlled $R_x(\theta)$ noise into each path's circuit after the entangling gates. The noise profiles cross: Path~A starts clean ($\theta_A \approx 0.04$~rad) and degrades ($\theta_A \to 1.22$~rad), while Path~B starts noisy ($\theta_B \approx 0.89$~rad) and improves ($\theta_B \to 0.02$~rad), crossing at approximately cycle~4. This models the common scenario where a previously reliable qubit chain degrades due to drift while an alternative path improves.

\subsubsection{Results}
\begin{figure}[t]
    \centering
    \includegraphics[width=\linewidth]{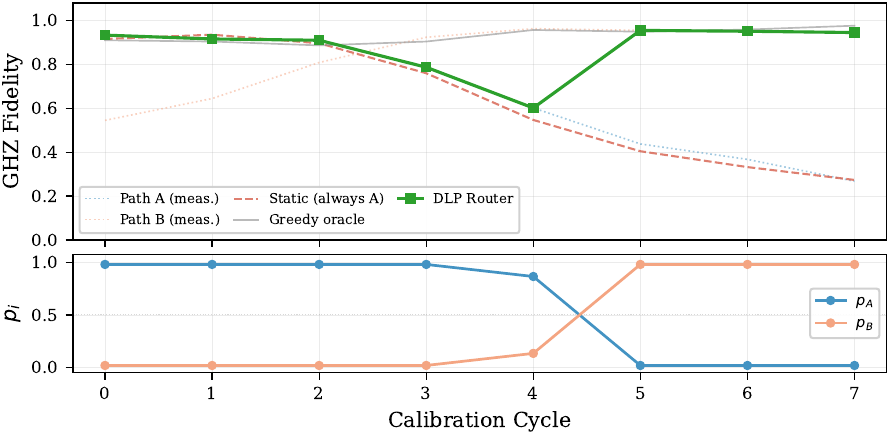}
    \caption{\textbf{Adaptive routing under gradual noise drift on \texttt{ibm\_fez}.} (Top)~GHZ fidelity comparison: the DLP router (green) tracks the better path, while the static baseline (red dashed) degrades with Path~A. (Bottom)~Softmax path probabilities showing the autonomous transition from $p_A \approx 1$ to $p_B \approx 1$ at the noise crossover.}
    \label{fig:crossing_combined}
\end{figure}

Figure~\ref{fig:crossing_combined} presents the results over 8 calibration cycles. The router locks onto Path~A immediately (cycles~0--3, $p_A > 0.98$), correctly identifying it as the higher-fidelity path. At cycle~4, the noise crossover is detected: the logits begin shifting ($\lambda_A = 0.94$, $\lambda_B = -0.94$), though the router still selects~A for one additional cycle, incurring a fidelity penalty (0.60 vs.\ the available 0.96 on Path~B). By cycle~5, the logits have fully flipped to $\lambda_B = +2.0$, and the router selects Path~B for the remaining cycles, achieving fidelities of 0.96--0.95.

The one-cycle delay at the crossover point is a direct consequence of the softmax dynamics: the accumulated gradient evidence from prior cycles creates inertia that must be overcome before the logits can reverse. This is distinct from a greedy oracle, which would switch instantly. The DLP router averages $0.875$ fidelity across all 8 cycles, representing a $+24.2$pp improvement over the static baseline ($0.633$), which remains locked to the degrading Path~A.

\subsubsection{Experiment 6b: Catastrophic Hardware Failure}
\label{sec:catastrophic}

\paragraph{Methodology.}
To test resilience to sudden failure, we designed an unbiased experimental protocol that avoids any hardwired path preference. Before the experiment, a calibration phase measures both paths 5 times each with zero injected noise to determine which path the hardware naturally prefers. On \texttt{ibm\_fez}, calibration found Path~B to be superior ($F_B = 0.975 \pm 0.005$ vs.\ $F_A = 0.923 \pm 0.018$). The experiment then injects catastrophic $R_x(1.2)$ noise exclusively on the calibrated preferred path (Path~B) starting at cycle~5, while injecting zero noise on both paths during the normal phase (cycles~0--4). The static baseline always uses the calibrated preferred path, ensuring a fair comparison. The router logits are initialized to zero.

\begin{figure}[t]
    \centering
    \includegraphics[width=\linewidth]{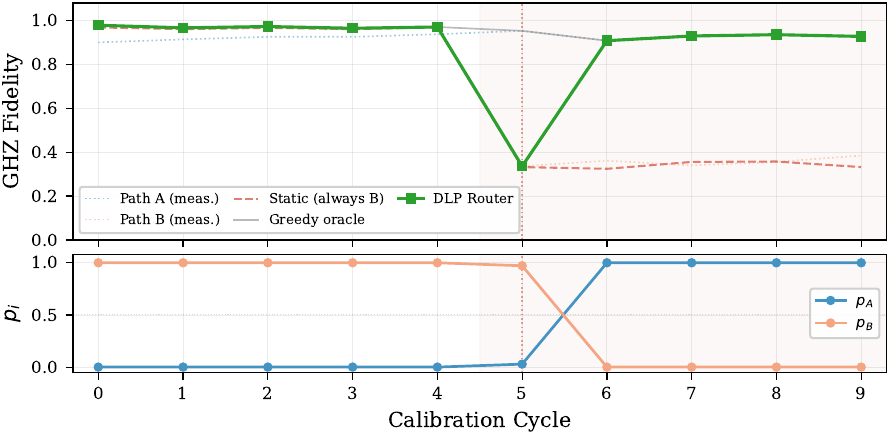}
    \caption{\textbf{Catastrophic failure recovery on \texttt{ibm\_fez}.} (Top)~GHZ fidelity: the DLP router (green) suffers one cycle of degradation at the failure onset (cycle~5), then recovers to ${\sim}0.93$ by rerouting to the survivor path. The static baseline (red dashed) remains stuck on the failed path. The shaded region indicates the failure phase. (Bottom)~Path probabilities showing the rapid logit reversal from $p_B \approx 1$ to $p_A \approx 1$ within one cycle of the failure.}
    \label{fig:catastrophic_combined}
\end{figure}

\paragraph{Results.}
Figure~\ref{fig:catastrophic_combined} shows the results over 10 cycles. During the normal phase (cycles~0--4), the router organically discovers Path~B as the better path, converging to $p_B > 0.99$ within the first cycle and achieving an average fidelity of $0.971$---comparable to the static baseline ($0.965$). No information about which path is better was provided to the router; this preference emerged purely from measurement-driven gradient updates.

At cycle~5, the injected failure destroys Path~B fidelity ($F_B = 0.34$). Critically, the DLP router is still committed to Path~B at this point and suffers the same catastrophic hit as the static baseline ($F_{\text{DLP}} = 0.34$, $F_{\text{static}} = 0.33$). This confirms that the router has no oracle knowledge of the failure.

The recovery is immediate: the large fidelity gradient at cycle~5 shifts the logits from $(\lambda_A, \lambda_B) = (-3.0, +3.0)$ to $(-1.7, +1.7)$, and by cycle~6 the logits have fully reversed to $(+3.0, -3.0)$, selecting Path~A with $F = 0.91$. The post-failure average fidelity is $0.807$ for DLP versus $0.340$ for the static baseline, a +46.7pp improvement. Over all 10 cycles, DLP achieves $0.889$ average fidelity compared to $0.653$ for static ($+23.6$~pp).

\begin{table}[h]
    \centering
    \caption{\textbf{Hardware experiment summary on \texttt{ibm\_fez}.} Average GHZ fidelity for DLP and baselines across both routing experiments.}
    \label{tab:hw_summary}
    \resizebox{1.0\linewidth}{!}{
    \begin{tabular}{lcccc}
        \toprule
        \textbf{Experiment} & \textbf{DLP} & \textbf{Static} & \textbf{Greedy} & \textbf{DLP vs Static} \\
        \midrule
        Crossing (8 cycles)     & 0.875 & 0.633 & 0.931 & $+24.2$~pp \\
        Catastrophic (10 cycles) & 0.889 & 0.653 & 0.951 & $+23.6$~pp \\
        \quad Post-failure only  & 0.807 & 0.340 & ---   & $+46.7$~pp \\
        \bottomrule
    \end{tabular}}
\end{table}

These results provide empirical evidence that differentiable logic programming can function as a form of \emph{online adaptive compilation}, continuously monitoring realized performance via quantum measurements and adjusting circuit structure accordingly---a capability that current rule-based compilers cannot provide. The crossing experiment demonstrates adaptation to gradual drift, while the catastrophic experiment demonstrates rapid recovery from sudden failure, with the unbiased calibration protocol ensuring that no prior path knowledge contaminates the result.

\section{Conclusion}
\label{sec:conclusion}

We have introduced a Differentiable Logical Programming (DLP) framework for quantum circuit design, a neuro-symbolic approach that recasts the NP-hard problem of circuit optimization into a continuous, gradient-based search. By representing a scaffold of candidate gates as learnable switches $\boldsymbol{s}$ (derived from structural logits $\boldsymbol{\lambda}$) and optimizing them to satisfy differentiable logical axioms, our method provides a uniquely flexible tool for algorithm design. We have demonstrated that this single, unified framework can: (1) function as a ``smarter'' compiler discovering non-trivial optimizations from first principles; (2) perform circuit discovery, such as the 12-gate 4-qubit QFT; (3) solve joint structural-parametric design problems; (4) scale to larger systems via Hierarchical Synthesis; and (5) perform fast hardware-aware adaptation to temporal noise and failures on the 156-qubit IBM Fez processor, including a +46.7~pp post-failure fidelity recovery through purely measurement-driven gradient updates.

While currently focused on scaffold pruning within a fixed gate ordering, future work will extend this to full topological search by incorporating differentiable ``Swap Networks.'' The flexibility of this approach, built on standard automatic differentiation libraries, allows for a diverse range of tasks simply by composing different logical axioms, providing a potent methodological response to the challenges of the NISQ and early fault-tolerant eras.

\subsection{Limitations and Scope: Fixed-Topology Optimization}
It is important to distinguish between \textit{topological discovery} (finding the optimal connectivity graph) and \textit{scaffold pruning} (optimizing a selection from an ordered list). In this work, we focus on the latter. The DLP framework is currently constrained by the fixed ordering of the input scaffold $S$. If the optimal solution requires a permutation of gates not present in $S$, the model cannot discover it. However, this constraint is often intentional for the use case of \textit{optimizing existing ansatzes}, where the ordering is dictated by a template. Future work will extend this to full topological search by incorporating differentiable ``Swap Networks.''

The primary strength of this approach is its flexibility and ease of implementation. The same core model, built on standard \texttt{autograd} libraries like PyTorch, can solve this diverse range of tasks simply by composing different logical axioms in the loss function.

\subsection{Data and Code Availability}
The source code and data supporting the findings of this study are available at \url{https://github.com/sulcantonin/dlp_public}. 

\subsection{Use of AI Tools}
The author used Gemini 2.5 Pro and Claude Sonnet as assistive tools during the writing and coding stages of this work, including text editing, code drafting, and debugging support.

\begin{acknowledgments}
This work was supported by the Director of the Office of Science of the U.S. Department of Energy under Contract No. DE-AC02-05CH11231.
\end{acknowledgments}

\bibliographystyle{quantum}
\bibliography{references}

\clearpage
\onecolumn
\appendix
\section*{Additional Differentiable Axioms}
\label{app:axioms}

Here we provide the mathematical forms for additional differentiable predicates $\mathcal{T}$ and their corresponding loss terms $\mathcal{L}$ that supplement the core axioms defined in Section~\ref{sec:axioms}.

\subsection*{Entanglement:}
Used for discovery of entangled states.
\begin{itemize}
    \item \textbf{Predicate $\mathcal{T}_{\text{ent}}$:} Based on the Von Neumann entropy $S(\rho_A)$ of the reduced density matrix $\rho_A$ for a bipartite split $A|B$.
    \begin{equation}
        \mathcal{T}_{\text{ent}}(U) = 1 - \exp(-k S(\rho_A(U))),
    \end{equation}
    where $S(\rho_A) = -\text{Tr}(\rho_A \log_2 \rho_A)$ and $k$ is a scaling factor.
    \item \textbf{Contradiction $\mathcal{L}_{\text{ent}}$:}
    \begin{equation}
        \mathcal{L}_{\text{ent}} = 1 - \mathcal{T}_{\text{ent}} = \exp(-k S(\rho_A)).
    \end{equation}
    This loss term is $\approx 1$ for separable states ($S=0$) and $\to 0$ for maximally entangled states.
\end{itemize}

\subsection*{Robustness (Noise):}
Used for multi-objective, noise-aware optimization.
\begin{itemize}
    \item \textbf{Predicate $E_{\text{noisy}}$:} The average energy under a set of error channels $\mathcal{N} = \{N_j\}$, representing a noise model.
    \begin{equation}
        E_{\text{noisy}} = \frac{1}{|\mathcal{N}|} \sum_{N_j \in \mathcal{N}} \langle \psi | N_j^\dagger H N_j | \psi \rangle,
    \end{equation}
    where $|\psi\rangle = U|0\rangle$.
    \item \textbf{Contradiction $\mathcal{L}_{\text{rob}}$:}
    \begin{equation}
        \mathcal{L}_{\text{rob}} = E_{\text{noisy}}.
    \end{equation}
    This loss term is used in a multi-objective function, e.g., $\mathcal{L}_{\text{total}} = w_E \mathcal{L}_{\text{energy}} + w_R \mathcal{L}_{\text{rob}} + w_S \mathcal{L}_{\text{simp}}$.
\end{itemize}

\clearpage
\section*{Supplementary Experiments}
\label{app:supplementary_experiments}

\subsection{Robust VQE Ansatz Discovery with Annealing Curriculum}
\label{app:vqe_annealing}

We applied the DLP framework to the nontrivial problem of finding the ground state for a 4-qubit 1D Ising model. This task is a common benchmark for Variational Quantum Eigensolvers (VQE) because the energy landscape contains a local minimum at $E = -4.0$ (a simple product state) that can trap gradient-based optimizers, preventing them from reaching the entangled ground state at $E \approx -4.73$. Our goal was to show that the framework can search this energy landscape and identify a simpler, noise-aware ansatz structure by pruning unnecessary gates from a generic hardware-efficient template.

\begin{figure}[t]
    \centering
    \begin{tabular}{c}
    \resizebox{1.0\linewidth}{!}{
    \begin{tabular}{cc}
    (a) Input Scaffold & (b) Optimized Scaffold ($\sigma=0.1$)\\
    \includegraphics[width=1.0\linewidth]{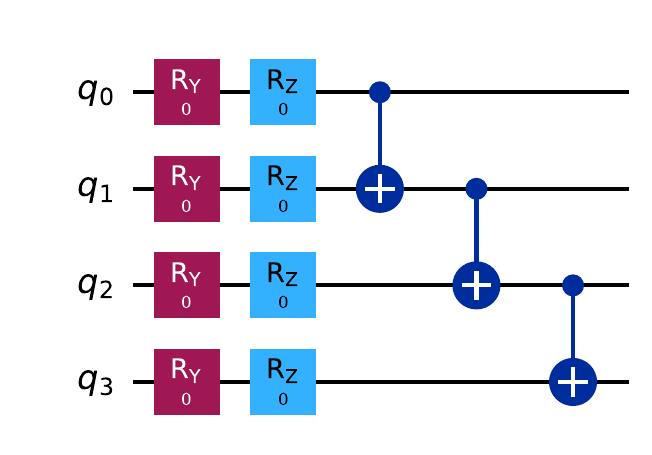} & 
    \includegraphics[width=0.7\linewidth]{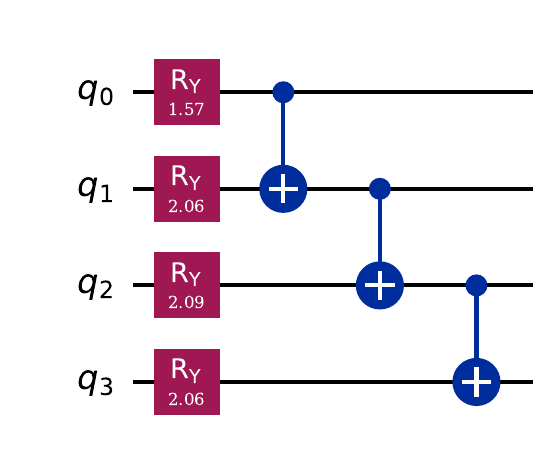}
    \end{tabular}}\\
    \includegraphics[width=1.0\linewidth]{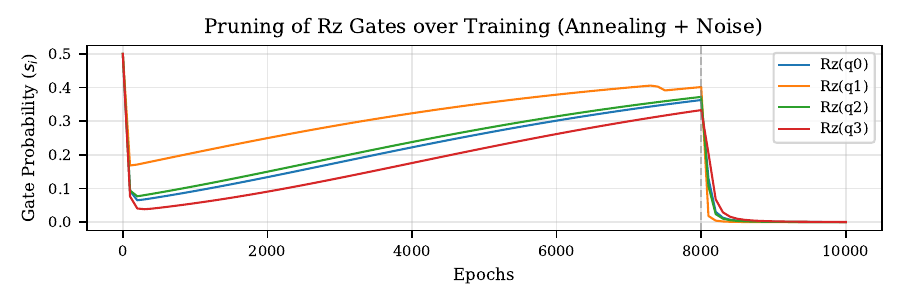}
    \end{tabular}
    \caption{\textbf{VQE Discovery Results.} \textbf{(a)} The input scaffold represents a standard, over-parameterized ansatz. \textbf{(b)} The final circuit structure autonomously discovered by the model. Note that the $R_z$ gates have been removed, leaving a minimal structure of $R_y$ rotations and entangling $CNOT$s. \textbf{(c)} The evolution of the gate switch probabilities ($s_i$) for the $R_z$ gates. As the curriculum introduces the interaction term $J$ (epochs $0-8000$) and the simplicity weight $w_{\text{simp}}$ increases (epoch $8000+$), the model decisively prunes these redundant operations, converging to a sparse topology while maintaining the ground state energy.}
    \label{fig:vqe_results}
\end{figure}

The experiment utilized a scaffold composed of a standard hardware-efficient ansatz (HEA) layer, including parameterized $R_y(\boldsymbol{\theta})$ and $R_z(\boldsymbol{\phi})$ rotation gates and a chain of entangling $CNOT$ operations. To overcome the local minima problem, we employed the \textit{Annealing Curriculum} strategy (Section~\ref{sec:curricula}). Instead of optimizing for the full Hamiltonian $H$ immediately, we trained the model on a time-dependent Hamiltonian $H(t)$, where the interaction term $J$ was slowly ramped from $0$ to $1.0$ over $8,000$ epochs, see Figure~\ref{fig:vqe_results} (c). 

To further test robustness, we injected Gaussian noise ($\sigma=0.1$) into the energy evaluation at every step. The optimization was driven by a composite loss function $\mathcal{L} = \mathcal{L}_{\text{energy}} + w_{\text{simp}}\mathcal{L}_{\text{simp}}$, which balanced the minimization of energy with a penalty for circuit complexity.

\paragraph{Results.} As the annealing schedule progressed (epochs $0-8000$), the system successfully escaped the $E=-4.0$ trap, converging to an energy of $E \approx -4.83$, which fluctuates around the theoretical ground state of $-4.73$ due to the injected noise. The structural learning (Fig.~\ref{fig:vqe_results} (c)) reveals a physically intuitive result: the optimizer systematically pruned the $R_z$ gates (driving their probabilities $s_i \to 0$) while retaining the $R_y$ and $CNOT$ gates. This indicates the model correctly identified that for this real-valued Hamiltonian, $Z$-rotations were redundant, thereby autonomously simplifying the ansatz to its most efficient form without human intervention.

\subsection{QAOA Depth Discovery}
\label{app:qaoa_depth}

While the main text focuses on circuit discovery and pruning in fixed scaffolds, this experiment isolates the framework's ability to tune a discrete hyperparameter: circuit depth. 

We tasked the model with finding the minimal QAOA (Quantum Approximate Optimization Algorithm) depth for a 4-node MaxCut instance, initializing a scaffold with $p_{\text{max}}=3$ layers, 
$$S=[U_C(\gamma_1),U_B(\beta_1),\dots,U_C(\gamma_3),U_B(\beta_3)].$$ 
Using a Two-Phase Curriculum: (1) we first optimized angles for energy with $w_{\text{simp}}=0$, and then  (2) applied simplicity pressure with $w_{\text{simp}}>0$ to encourage pruning. The optimizer converged to $s_i\approx 0$ for the final ($p=3$) layer, indicating that a reduced depth of $p=2$ is sufficient for this problem.

\end{document}